\documentclass[]{spie}  

\usepackage{amsmath,amsfonts,amssymb}
\usepackage{graphicx}
\usepackage[colorlinks=true, allcolors=blue]{hyperref}
\usepackage{mathrsfs}
\usepackage{float}

\title{Comparing Complex Impedance and Bias Step Measurements of Simons Observatory Transition Edge Sensors}

\author[a]{Nicholas F. Cothard}
\author[b]{Aamir M. Ali}
\author[c]{Jason E. Austermann}
\author[d,e]{Steve K. Choi}
\author[b]{Kevin T. Crowley}
\author[c]{Bradley J. Dober}
\author[d]{Cody J. Duell}
\author[c]{Shannon M. Duff}
\author[d]{Patricio Gallardo}
\author[c]{Gene C. Hilton}
\author[f]{Shuay-Pwu Patty Ho}
\author[c]{Johannes Hubmayr}
\author[c]{Micheal J. Link}
\author[d,e]{Michael D. Niemack}
\author[g]{Rita F. Sonka}
\author[g]{Suzanne T. Staggs}
\author[d]{Eve M. Vavagiakis}
\author[h]{Edward J. Wollack}
\author[i,j]{Zhilei Xu}

\affil[a]{Department of Applied and Engineering Physics, Cornell University, Ithaca, NY, USA}
\affil[b]{Department of Physics, University of California, Berkeley, Berkeley, CA, USA}
\affil[c]{Quantum Devices Group, NIST, Boulder, CO, USA}
\affil[d]{Department of Physics, Cornell University, Ithaca, NY, USA}
\affil[e]{Department of Astronomy, Cornell University, Ithaca, NY, USA}
\affil[f]{Department of Physics, Stanford University, Princeton, NJ, USA}
\affil[g]{Department of Physics, Princeton University, Princeton, NJ, USA}
\affil[h]{NASA Goddard Space Flight Center, Greenbelt, MD, USA}
\affil[i]{Department of Physics and Astronomy, University of Pennsylvania, Philadelphia, PA, USA}
\affil[j]{MIT Kavli Institute, Massachusetts Institute of Technology, Cambridge, MA, USA}

\authorinfo{Further author information: (Send correspondence to N.F.C.)\\N.F.C.: E-mail: nc467@cornell.edu, Telephone: 1 607 255 0833}

\begin{document} 
\maketitle

\begin{abstract}
The Simons Observatory (SO) will perform ground-based observations of the cosmic microwave background (CMB) with several small and large aperture telescopes, each outfitted with thousands to tens of thousands of superconducting aluminum manganese (AlMn) transition-edge sensor bolometers (TESs). In-situ characterization of TES responsivities and effective time constants will be required multiple times each observing-day for calibrating time-streams during CMB map-making. Effective time constants are typically estimated in the field by briefly applying small amplitude square-waves on top of the TES DC biases, and fitting exponential decays in the bolometer response. These so-called ``bias step'' measurements can be rapidly implemented across entire arrays and therefore are attractive because they take up little observing time. However, individual detector complex impedance measurements, while too slow to implement during observations, can provide a fuller picture of the TES model and a better understanding of its temporal response. Here, we present the results of dark TES characterization of many prototype SO bolometers and compare the effective thermal time constants measured via bias steps to those derived from complex impedance data.

\end{abstract}

\section{Introduction}
\label{sec:intro}
The Simons Observatory (SO) is a suite of multiple small ($\sim0.5$ m) and one large ($\sim6$ m) aperture telescopes (SATs and LAT, respectively) that will utilize superconducting transition-edge sensor (TES) bolometers to map the cosmic microwave background (CMB) \cite{galitzki_simons_2018}.
SO will be located at an altitude of 5190 m on Cerro Toco in the Atacama Desert. 
The combination of the dry site and the capability to measure multiple angular scales will enable many science goals using the same detector array technologies \cite{ade_simons_2019}.
For example, SO will improve constraints on cosmological parameters, probe the sum of the neutrino masses, detect high-redshift galaxy clusters, and use gravitational lensing to characterize the distribution of dark matter.

In order to achieve its broad science goals, SO will deploy $\sim70,000$ polarization sensitive AlMn TES detectors in six frequency bands spanning 30 -- 290 GHz. 
Three dichroic pixel designs will constitute the low-frequency (LF 30/40 GHz), mid-frequency (MF 90/150 GHz), and ultra-high-frequency (UHF 230/290 GHz) arrays.
The LF arrays are being fabricated at the University of California at Berkeley (UCB) and will be optically coupled via sinuous antennas and silicon lenslets.
The MF and UHF arrays are being fabricated at the National Institute of Standards and Technology at Boulder (NIST) and will be optically coupled via orthomode transducer antennas and gold-plated aluminum feedhorns.

The TES bolometer properties must be optimized for the expected observing and loading conditions.
Section \ref{sec:background} discusses the single thermal block model used here.
Given the loading at each frequency band, $P_\gamma$, the bolometer critical temperature, $T_\textrm{c}$ and thermal conductance to the bath, $G$, are tuned to achieve target saturation powers, $P_\textrm{sat}$  \cite{stevens_characterization_2020}.
Similarly, the bolometer heat capacity, $C$, is tuned to achieve the target time constants, $\tau_\textrm{eff}$, required for efficient sampling of the sky.
SO time constant specifications are driven by the telescope scanning and observing modes.
The SATs will utilize a broad-band, rotating, cryogenic half-wave plate (CHWP) to minimize polarization systematics by modulating the incident polarization.
In order to deconvolve the modulated signal from a 2 Hz CHWP with sufficient accuracy, SO places a minimum time constant requirement on detectors of $\tau_\textrm{eff} \leq 1.1$ ms or $f_\textrm{3dB}=1/2\pi\tau_\textrm{eff} \geq 150$ Hz.
The time constant requirement for detectors in the LAT are driven by the combination of the telescope's diffraction limited beam size and telescope scan speed.
Table \ref{tab:requirements} gives the target saturation power ranges and time constants for each frequency band.
 
\begin{table}[ht]
\caption{Target saturation power ($P_\textrm{sat}$) and slowest time constants ($f_\textrm{3dB,min}$) for each SO detector band. The time constant specification of the three lowest frequency bands is driven by the requirement to deconvolve the CHWP signal. The time constant specification of the three highest frequency bands is driven by the LAT beam size and scan speed.} 
\label{tab:requirements}
\begin{center}       
\begin{tabular}{|c|c|c|}
\hline
\rule[-1ex]{0pt}{3.5ex}  Frequency Band & Target $P_\textrm{sat}$ Range & Target $f_\textrm{3dB,min}$  \\
\hline
\rule[-1ex]{0pt}{3.5ex}  LF-1 30 GHz & 0.6 -- 1.0 pW & 150 Hz   \\
\hline
\rule[-1ex]{0pt}{3.5ex}  LF-2 40 GHz & 2.7 -- 4.4 pW & 150 Hz   \\
\hline 
\rule[-1ex]{0pt}{3.5ex}  MF-1 90 GHz & 2.0 -- 3.3 pW & 150 Hz   \\
\hline
\rule[-1ex]{0pt}{3.5ex}  MF-2 150 GHz & 5.4 -- 9.0 pW & 166 Hz   \\
\hline
\rule[-1ex]{0pt}{3.5ex}  UHF-1 230 GHz & 16.9 -- 28.1 pW & 245 Hz   \\
\hline
\rule[-1ex]{0pt}{3.5ex}  UHF-2 290 GHz & 22.4 -- 37.3 pW & 279 Hz   \\
\hline
\end{tabular}
\end{center}
\end{table}

In this paper, we focus on measurements of bolometer time constants by electro-thermal stimulation.
A commonly used diagnostic for fast, in-situ, array-scale time constant measurements is the bias step (BS) method (Section \ref{sec:bs}).
In this method, each bias line in an array is modulated with a small amplitude square wave, and each detector’s time constant is extracted from the exponential decay in the bolometer responses.
This is ideal for array characterization in the field since it can be measured on all detectors on a given bias line simultaneously, minimizing the diagnostic time.
An alternative diagnostic approach is to extract the complex impedance (CZ) of each bolometer by measuring their response to sinusoidal stimuli at a variety of frequencies (Section \ref{sec:cz}). 
With an electrothermal bolometer model, CZ measurements enable deeper probes into the device physics of the detectors.
However, with existing readout systems, CZ measurements are much slower than BS measurements due to the need to sample each detector very quickly (limiting the multiplexing rate), making them unsuitable for fast in-situ array-scale characterization.
Here, we use both the BS and CZ methods to measure the time constants of SO prototype single pixel detectors and then compare the results of both methods.

\section{Background}
\label{sec:background}
\begin{figure} [ht]
\begin{center}
\begin{tabular}{cc}
\includegraphics[height=6cm]{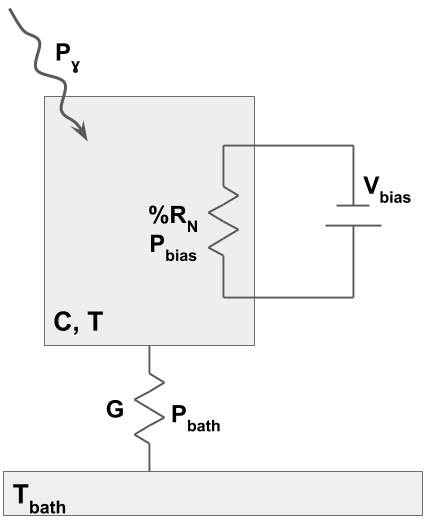} & \includegraphics[height=5cm]{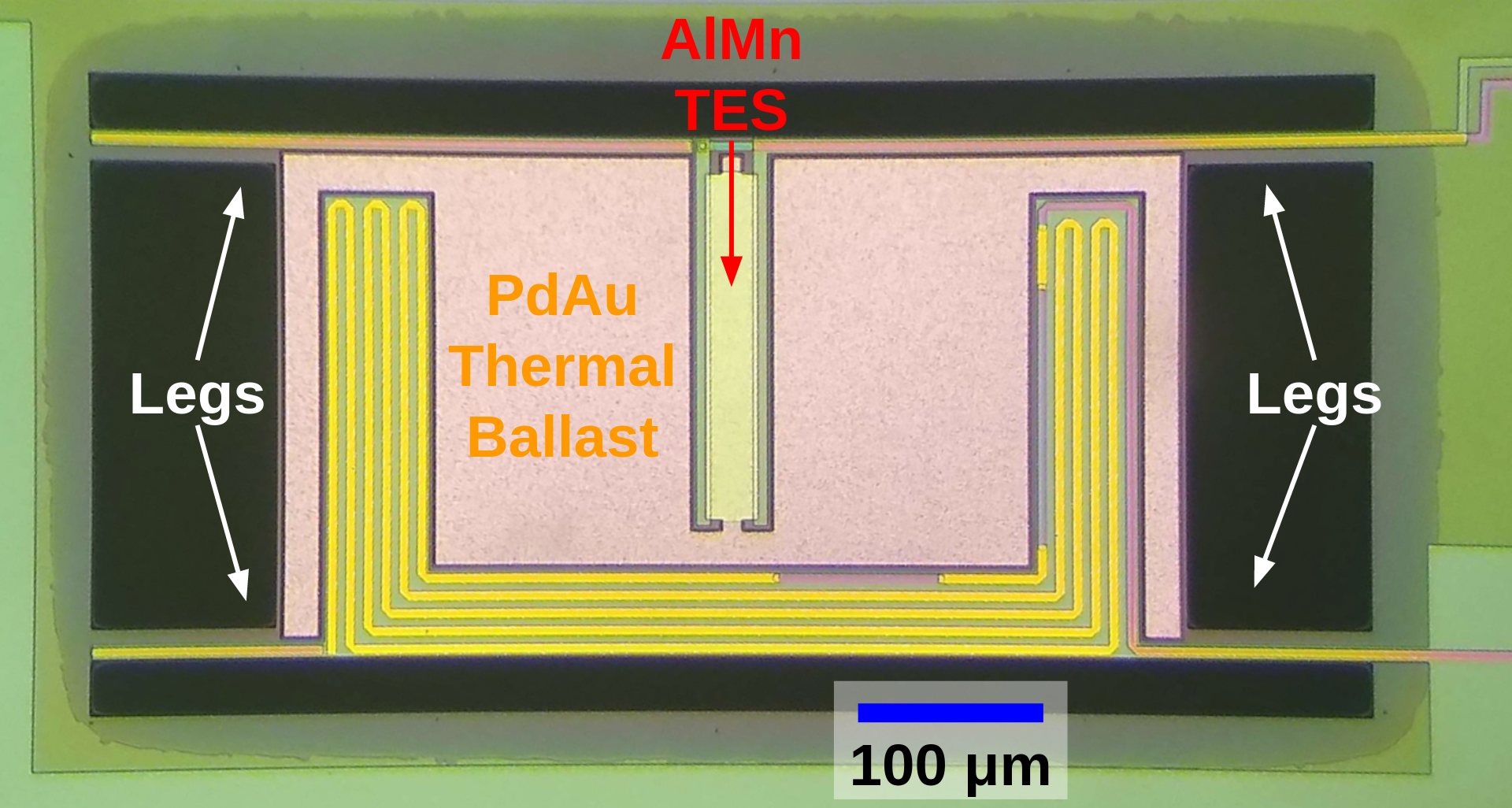}
\end{tabular}
\end{center}
\caption
{\label{fig:bolo} 
\emph{Left}: Diagram of the single bolometer island model used here including a single heat capacity, $C$, and thermal conductance to the bath, $G$. \emph{Right}: Microscope photograph of a prototype SO 150 GHz TES bolometer fabricated at NIST. Four legs support the floating island and define $G$. The meandering line in the lower portion of the island is the resistive termination from the waveguide. The AlMn TES is the off-white rectangle in the middle of the island. The surrounding pink area consists of extra AlMn and the PdAu thermal ballast. 
}
\end{figure}

We use a simple single thermal block model for the bolometer as is well developed in Irwin and Hilton’s Transition-Edge Sensors \cite{irwin_transition-edge_2005}.
In this model, a TES bolometer consists of a single lumped heat capacity, $C$, at temperature, $T$, and with a thermal conductance, $G$, to the bath with temperature, $T_\textrm{bath}$. 
On the bolometer island, the TES is voltage biased to some percentage of its normal resistance, $\%R_\textrm{N}$.
Figure \ref{fig:bolo} \emph{left} shows a schematic of this model including the incident power from radiation, bias power, and the power flowing to the bath $P_\textrm{bath} = P_\gamma + P_\textrm{bias}$. 
Figure \ref{fig:bolo} \emph{right} shows a photo of an SO 150 GHz TES bolometer island fabricated at NIST. 
The TES is suspended on a thin film via four “legs” whose width and length control the thermal conductance to the bath.
The heat capacity is controlled by adjusting the volume of thermal ballast (PdAu) on the suspended island. 
The intrinsic time constant (in the absence of electrical feedback, $P_\textrm{bias}=0$) of the bolometer is therefore controlled by geometric properties of the bolometer island, $\tau_\textrm{nat} = C/G$.

In the small signal limit, when a voltage bias is applied, negative electrothermal feedback can be achieved to keep the TES on transition \cite{irwin_transition-edge_2005}. 
The coupled electrothermal system gives rise to electrical and effective thermal time constants. 
The electrical time constant, $\tau_\textrm{el}$, is determined by TES bias circuit elements and the TES current sensitivity, $\beta = \frac{\textrm{d}\log{R}}{\textrm{d}\log{I}}$.
The effective thermal time constant, $\tau_\textrm{eff}$, is slower than the $\tau_\textrm{el}$ and is our primary quantity of interest when considering the bolometer’s response to changes in loading from the sky. 
Following Irwin and Hilton, we write the effective thermal time constant as
\begin{equation}
f_\textrm{3dB} = f_\textrm{nat}\left(1 + \frac{\mathscr{L}}{1+\beta} \right) = 
\frac{G}{2\pi C}\left(1+\frac{1}{(1+\beta)}\frac{\alpha P_\textrm{bias}}{T_\textrm{c}G}\right)
\label{eq:f3db}
\end{equation}
where $f_\textrm{nat} = 1/2\pi\tau_\textrm{nat} = G/2\pi C$ is the natural thermal time constant, $\mathscr{L}=\alpha P_\textrm{bias}/T_\textrm{c}G$ is the loop gain, and $\alpha=\frac{\textrm{d}\log{R}}{\textrm{d}\log{T}}$ is the TES sensitivity to small changes in temperature \cite{irwin_transition-edge_2005}.
In this model, we see that the effective thermal time constant is proportional to the natural time constant, fundamental TES properties, and the bias power.
In practice, when optimizing a bolometer for on-sky performance, $G$ is first tuned to satisfy the saturation power requirements and then the time constant is tuned by changing $C$.
Single pixel fabrication runs are used to characterize different prototype bolometers with multiple variations of leg lengths and thermal ballasts.

The fundamental TES and bolometer parameters can be accessed through the complex impedance of the bolometer.
Irwin and Hilton derive the complex impedance such that
\begin{equation}
Z_\textrm{TES}(f) = R(1+\beta) + \frac{R(2 +\beta) \mathscr{L}}{(1 -\mathscr{L}) +if/f_\textrm{nat}}
= R(1+\beta) + \frac{R(2 +\beta) \alpha P_\textrm{bias}}{(T_\textrm{c}G -\alpha P_\textrm{bias}) +2\pi ifT_\textrm{c}C}
\label{eq:ZTES}
\end{equation}
where $R$ is the operating resistance of the TES and $f$ is the excitation frequency.
Thus, if $Z_\textrm{TES}(f)$ can be measured at a variety of excitation frequencies and at multiple bias powers, the parameters $\alpha$, $\beta$, and $C$ can be fit.
The parameters $R_\textrm{N}$, $T_\textrm{c}$, $G$, and $P_\textrm{bias}$ are all determined via current-voltage (I-V) curve measurements prior to the CZ measurements.
$Z_\textrm{TES}(f)$ takes the form of a semi-circle in the lower half of the complex plane. 
The low frequency limit reduces to $Z_\textrm{TES} \approx - R$ where as the high frequency limit goes to $Z_\textrm{TES} \approx R(1+\beta)$.
With high-fidelity data at both high and low frequencies, degeneracies between $\beta$ and the other fit parameters can be minimized. 
Successful fits result in a more complete picture of TES parameters including the heat capacity and effective thermal time constant.

\section{Bias Step Measurements}
\label{sec:bs}
This work focuses on SO prototype bolometer characterization being carried out in a dark dilution refrigerator at Cornell University with a time division multiplexing (TDM) multichannel electronics (MCE) readout system \cite{battistelli_functional_2008}. 
I-V curve measurements of each bolometer at multiple bath temperatures are used to characterize $T_\textrm{c}$, $R_\textrm{N}$, $P_\textrm{sat}$, $G$, and $P_\textrm{bias}$.
Time constants are then estimated from bias step measurements at multiple points on the superconducting  transition.
During a bias step measurement using the MCE, a TES is biased onto its transition to some fraction of its normal resistance, $\%R_\textrm{N}$, and then a square wave with an amplitude much smaller ($\sim 5\%$) than $P_\textrm{bias}$ is applied on top of the DC bias.
Data were also taken with even smaller amplitude square waves and were found to be consistent with these results and so, we use the larger amplitude to increase the signal to noise for our time-domain fits.
Roughly 30 periods of the square waves are repeated in order to build a statistical picture of the bolometer’s stimulated response.
To understand how the time constant varies at different points on the superconducting transition, the TES can then be re-biased to a different $\%R_\textrm{N}$ and another bias step can acquired. 
The bias step time constant measurements reported here are measured with an MCE sampling rate of $\sim\!6.4$ kHz in order to fully resolve the thermal decay.

\begin{figure}[ht]
\begin{center}
\includegraphics[width=\textwidth]{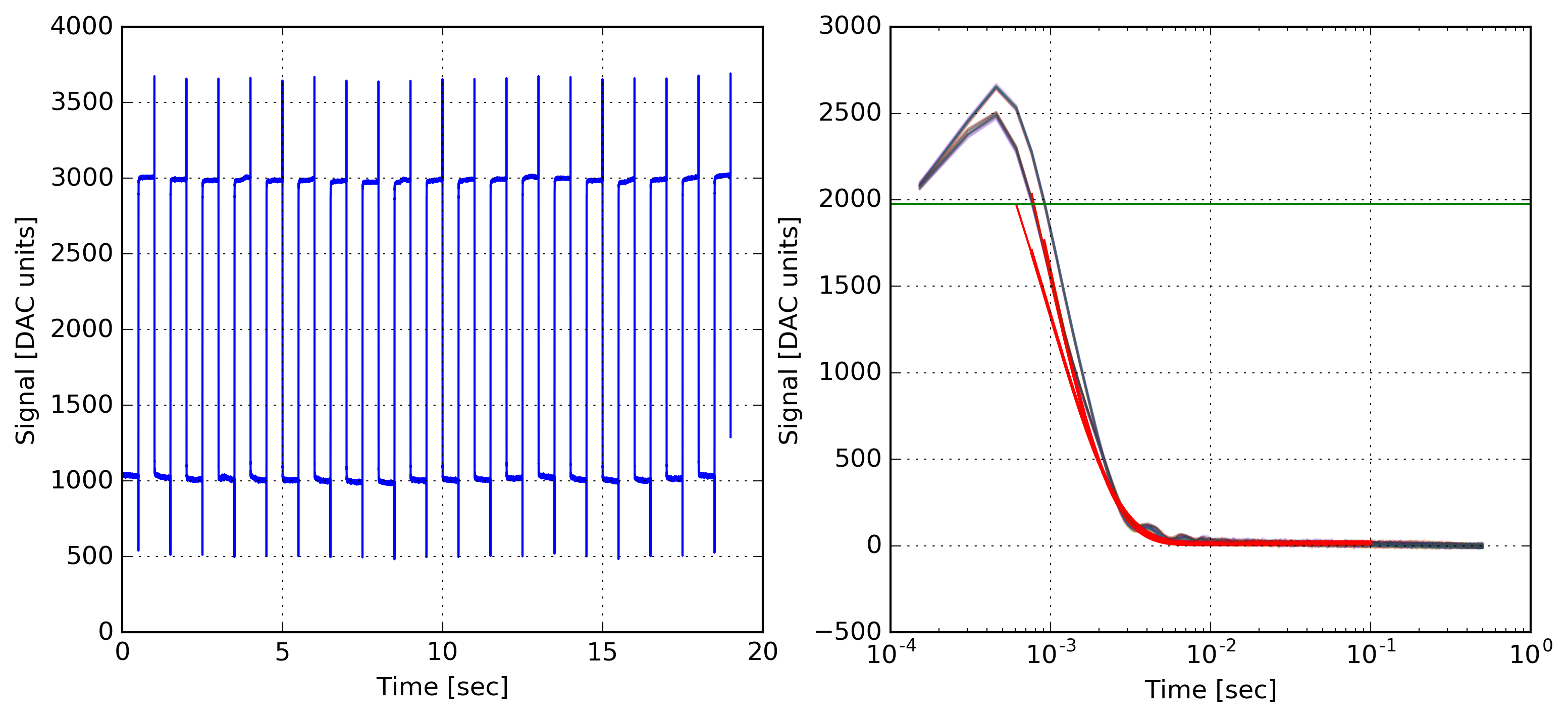}
\end{center}
\caption
{\label{fig:BS_timestream} 
\emph{Left}: Example bias step time stream for a prototype SO 150 GHz bolometer. The square wave response is clearly seen. The spikes before each step is the electrical response of the TES circuit. \emph{Right}: Each step is extracted, flipped, over plotted (colored lines), and fit (red lines). The horizontal green line indicates the height of the previous step. Data above the green line are excluded from the single pole exponential fit. 
}
\end{figure}

Before fitting the TES response, the time stream is split into individual steps of the square wave.
Each step is then fit to a single pole exponential of the form $f(t) = Ae^{-t/\tau}+C$.
In order to primarily fit the thermal response and not the electrical response, care is taken to choose the starting point of the fit. 
The exponential decay fit begins at the asymptotic value of the previous step.
Fitting prior to this would include the electrical response and would no longer fit well to a single pole filter.
Figure \ref{fig:BS_timestream} \emph{left} shows an example bias step time stream in DAC units. 
Each step (colored lines) and fit (red lines) are over plotted in Figure \ref{fig:BS_timestream} \emph{right} after being split from the time stream, flipped such that the step is always downwards, and given an vertical offset such that the asymptotic value of each step is zero. For fast detectors, the exponential decay fits will generally slightly under estimate the speed of the detectors response. This is due to the limited amount of data on the decay and can be fixed by sampling faster the detector faster. This effect can be seen in Figure \ref{fig:BS_timestream} \emph{right} as the red fits appear slightly slower than the colored data.

\begin{figure}[ht]
\begin{center}
\includegraphics[height=9cm]{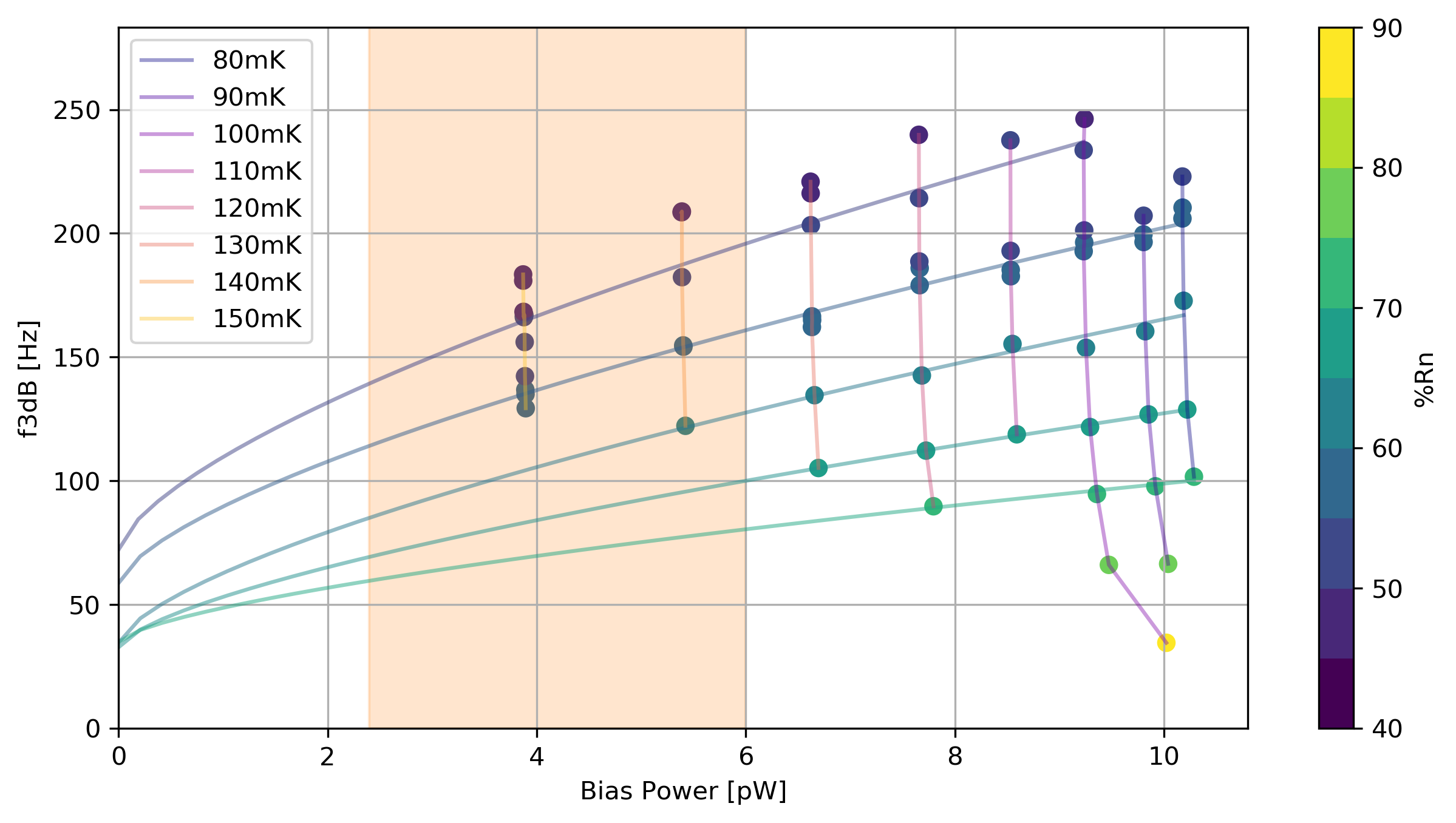}
\end{center}
\caption
{\label{fig:BS_pbias} 
Bias step time constants as a function of bias power and $\%R_\textrm{N}$ at many bath temperatures for a prototype SO 150 GHz bolometer. The vertical orange band shows the expected bias power range for this band when on-sky. We see that at $\sim 50 \%R_\textrm{N}$, we achieve the 166 Hz time constant specification of this band. over a broad range of bias powers. The upwards sloping lines fit the data to a two-fluid model \cite{irwin_thermal-response_1998} for fixed $\%R_\textrm{N}$. The fits are extrapolated to $P_\textrm{bias}=0$ to estimate the natural time constant which for this detector is between 40 -- 75 Hz. 
}
\end{figure} 

Time constants at different sky loading conditions can be simulated by measuring bias steps at multiple bath temperatures.
Figure \ref{fig:BS_pbias} shows time constants in $f_\textrm{3dB}$ of an example bolometer as a function of bias power and fraction of normal resistance. 
Each vertical grouping of data corresponds to a set of bias step measurements at one bath temperature.
As bath temperature increases, $P_\textrm{bias}$ decreases because $P_\textrm{sat}$ decreases, simulating optical loading.
The natural time constant of the bolometer can be estimated by extrapolating these results to $P_\textrm{bias}=0$, where the bolometer behavior is entirely thermal. 
A linear extrapolation would fit the simple one-block model described above but recent bias step time constant measurements \cite{koopman_advanced_2018} have found a two-fluid model \cite{irwin_thermal-response_1998} to fit better.
In this model, $f_\textrm{3dB}(P_\textrm{bias}) = A+BP_\textrm{bias}^{2/3}$.
Figure \ref{fig:BS_pbias} shows these extrapolations for fixed $\%R_\textrm{N}$.

\section{Complex Impedance Measurements}
\label{sec:cz}
To measure the complex impedance of the bolometers, small amplitude ($< 1\%$) sine waves are applied on top of the DC operating bias point determined from IV curves.
The bolometer transfer function is obtained by mapping the TES response as a function of stimulated frequency.
The MCE's arbitrary waveform generator (AWG) is used to apply digitized sine waves to the DC biases. 
The frequency at which the AWG can update the bias registers is related to the readout frequency and mode. 
Faster sampling modes enable faster excitations. 
These data were sampled at $\sim\!7.8$ kHz and with sine excitations ranging from 4 Hz to $\sim\!1.3$ kHz. 
  
\begin{figure}[ht]
\begin{center}
\includegraphics[width=\textwidth]{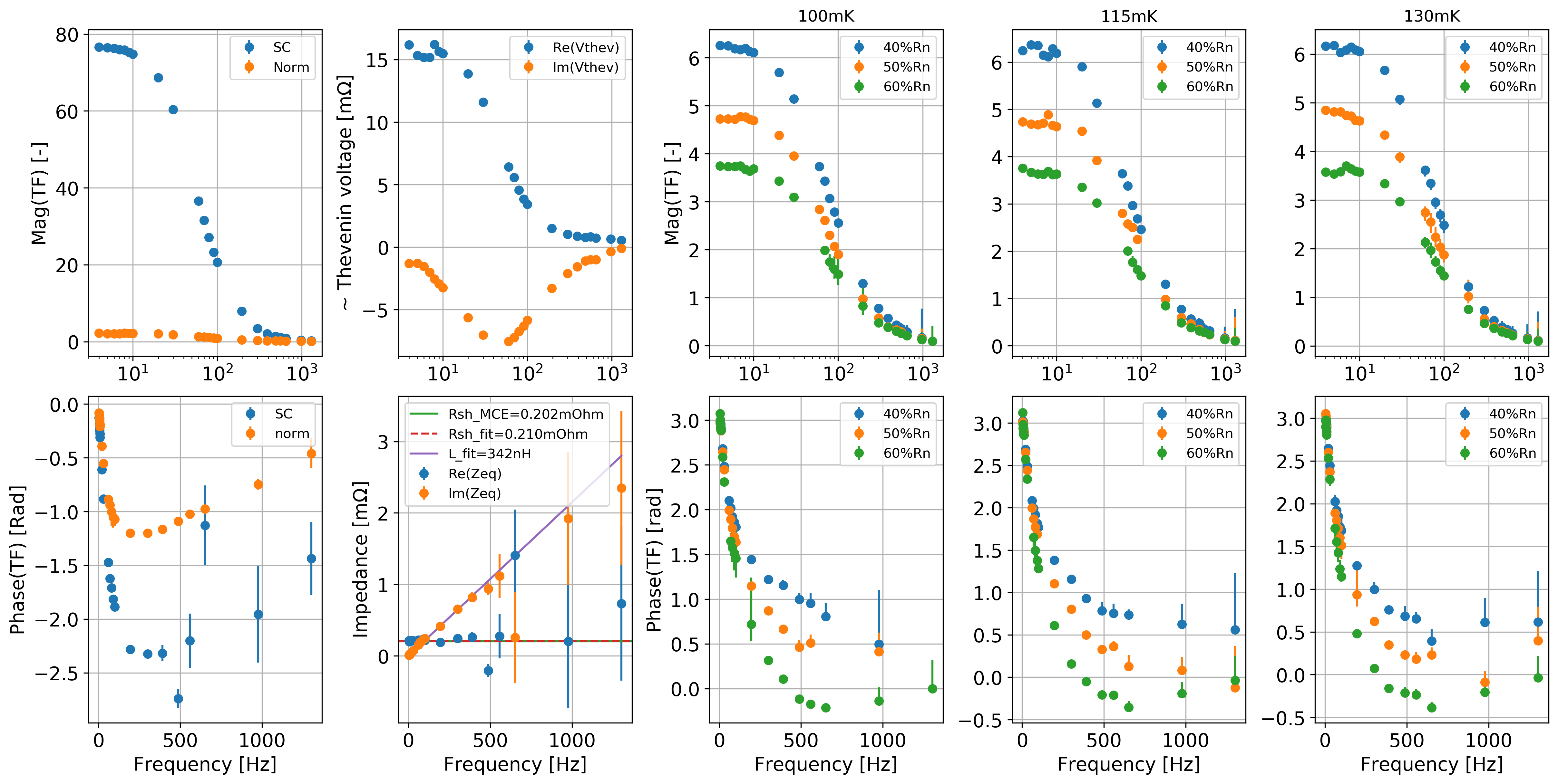}
\end{center}
\caption
{\label{fig:TF_ZEQ} 
Transfer functions of an example prototype SO 150 GHz bolometer showing the relative amplitude and phase of a TES bolometer's response to an input stimulus as a function of stimulus frequency. The left two columns show the transfer functions of the bolometer in its superconducting and normal states, which are used to extract the Thevenin voltage and TES bias circuit's equivalent impedance. The equivalent impedance plot shows that it is dominated by the shunt resistor and parasitic line inductance, as expected. The right-most three columns show the transfer functions of the on-transition bolometer at three different bias points and three different bath temperatures.
}
\end{figure}

Each sinusoidal response is fit for amplitude and phase.
As a function of the applied frequency, the relative amplitude and phase between the measured response and the input function forms a complex-valued transfer function.
This transfer function incorporates the TES response as well as the TES bias circuitry response.
The Thevenin equivalent voltage, ${V}_\textrm{th}$, and equivalent impedance, $Z_\textrm{eq}$ of the TES bias circuit are used to extract only the TES response, $Z_\textrm{TES}$.
These calibration values are obtained by measuring the transfer functions in the normal and superconducting states of the TES \cite{crowley_thesis_2018}.
We expect that $Z_\textrm{eq}$ is dominated by the bias shunt resistor and parasitic line inductance in the superconducting wiring such that $Z_\textrm{eq} = R_\textrm{sh} + 2\pi ifL_\textrm{par}$.
The left two columns of Figure \ref{fig:TF_ZEQ} plot the superconducting and normal transfer functions, ${V}_\textrm{th}$, and $Z_\textrm{eq}$, and confirms our expectations for $Z_\textrm{eq}$.
After calibrating the transfer function with $V_\textrm{th}$ and $Z_\textrm{eq}$, we obtain $Z_\textrm{TES}$. 
Uncertainties in $Z_\textrm{TES}$ are propagated from the sinusoidal fits.
For a given bolometer, transfer functions and $Z_\textrm{TES}$ are constructed at multiple $\%R_\textrm{N}$ and multiple $T_\textrm{bath}$.
The right three columns of Figure \ref{fig:TF_ZEQ} plot the magnitude and phase on-transition transfer functions of a TES at three different points on the transition, 40, 50, and 60 $\%R_\textrm{N}$, and at three different bath temperatures, 100, 115, and 130 mK.

\begin{figure}[ht]
\begin{center}
\includegraphics[width=\textwidth]{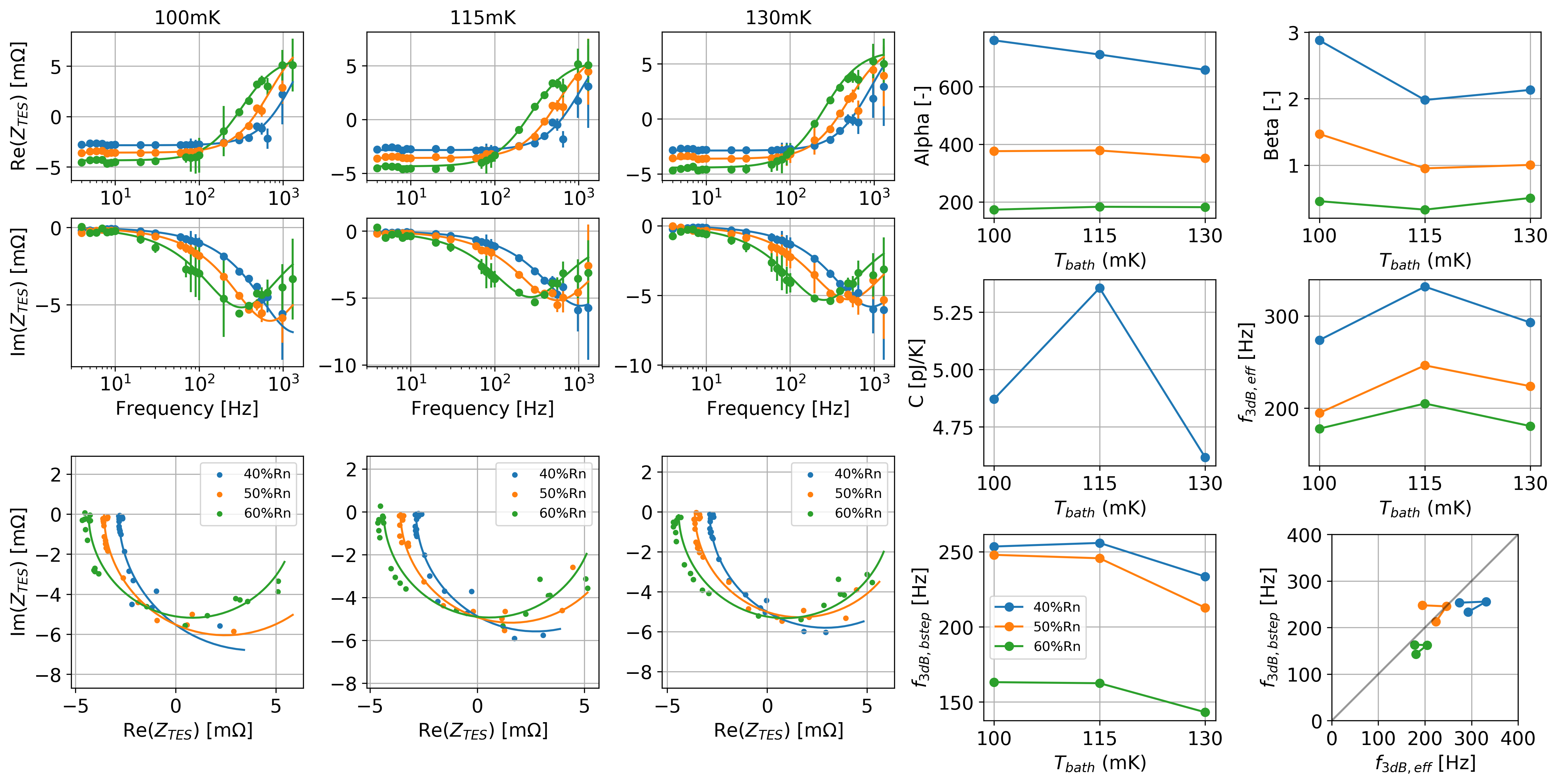}
\end{center}
\caption
{\label{fig:CZ_fit} 
Complex impedance and fits for an example prototype SO 150 GHz bolometer. Data and fits shown in the three left-most columns for each bias point and bath temperature. Fit results for $\alpha$, $\beta$, and $C$, as well as calculated values for the effective time constants shown in the two right most columns. Also shown are bias step measured effective time constants of the same bolometer at the same bias points and bath temperatures. A scatter plot shows that the complex impedance and bias step time constants are consistent.
}
\end{figure}

We now fit each $Z_\textrm{TES}$ measurement to the bolometer model described above in equation \ref{eq:ZTES} for $\alpha$, $\beta$, and $C$.
The heat capacity is assumed to be constant as a function of $\%R_\textrm{N}$ and therefore is constrained as such between all fits at the same $T_\textrm{bath}$.
I-V curve measurements of $R_\textrm{N}$, $T_\textrm{c}$, $G$, and $P_\textrm{bias}$ are used in the fitting and are assumed to have negligible uncertainties. 
A rough fit of the data is performed with a SciPy implementation of the Nelder-Mead algorithm and then improved using the MIGRAD algorithm \cite{crowley_thesis_2018,james_minuit_1975}.
Figure \ref{fig:CZ_fit} shows the complex impedance data and fits of an example bolometer at multiple $\%R_\textrm{N}$ and $T_\textrm{bath}$.
 
The left three columns of Figure \ref{fig:CZ_fit} plot the complex $Z_\textrm{TES}$ for each $\%R_\textrm{N}$ and bath temperature. 
Plotting the data on the complex plane demonstrates that the complex impedance of the bolometer follows the expected semicircular form implied by the single body model described by equation \ref{eq:ZTES}.
The fits to $Z_\textrm{TES}$ are over plotted as solid lines and also show that the data fits well to the model.
The right-most two columns show the fitted values for $\alpha$, $\beta$, and $C$ as a function of bias point and bath temperature.
The effective time constant is then inferred from the fitted values and plotted as $f_\textrm{3dB, eff}$.
Also plotted are the time constants of the same device measured with the bias step method, $f_\textrm{3dB, bstep}$, at the same bias points and bath temperatures.
A scatter plot shows that the time constants measured via complex impedance are in rough agreement with those measured via bias steps.

\section{Results}
\label{sec:results}
\begin{figure}[ht]
\begin{center}
\includegraphics[height=8cm]{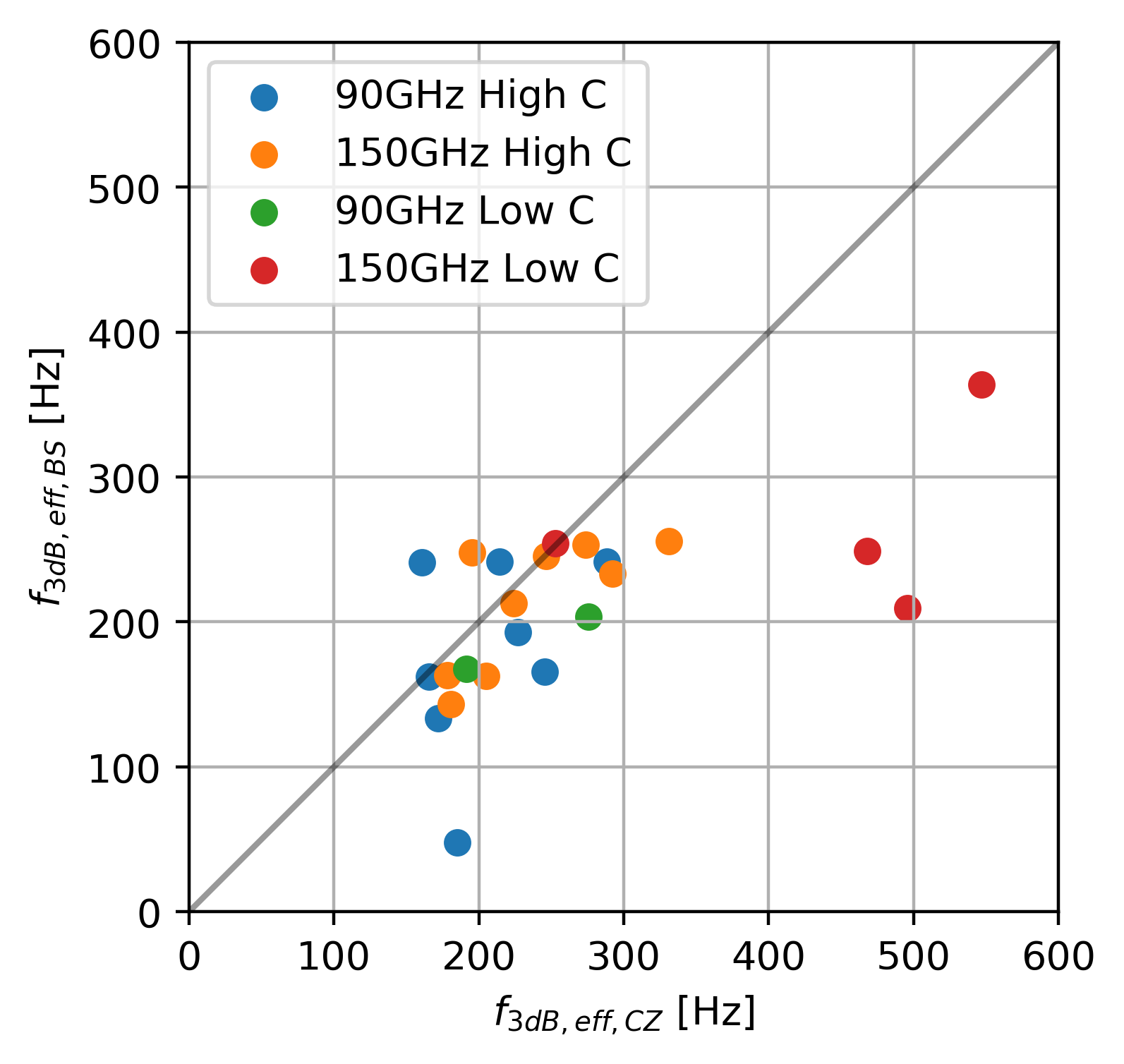}
\end{center}
\caption
{\label{fig:cz_vs_bs} 
Scatter plot of the effective thermal time constants determined via complex impedance and bias step measurements of four prototype SO MF bolometers. Low and high heat capacity bolometers were probed. The two methods are generally in agreement though the bias step method slightly underestimates the complex impedance method. Measurements of the fastest device with the smallest heat capacity and largest thermal conductance (red points) appear to diverge. This is likely due to insufficient high frequency excitations resulting in an incomplete mapping of this bolometer's response in the complex plane.
}
\end{figure}

Here we present a comparison of bias step and complex impedance measurements of the effective time constants of multiple prototype SO MF bolometers. 
The devices sampled include 90 and 150 GHz detectors with various amounts of heat capacities, controlled by the volume of thermal ballast on the island. 
These prototype bolometers were fabricated to explore the range of time constants achievable by tuning the bolometer heat capacities.
Figure \ref{fig:cz_vs_bs} directly compares the time constants of four devices determined via bias step and complex impedance measurements.
Each point on the scatter plot corresponds to a given bolometer (denoted by color) at a given bias point and bath temperature.

From these data, it is clear that while the two methods are in general agreement, the bias step data may slightly under estimate the complex impedance measurements.
The largest discrepancy occurs for the lowest heat capacity device (150 GHz Low C), which is expected to be the fastest and potentially least stable.
For this particular detector, as the heat capacity is lowered and the device is made faster, it becomes more challenging to fully probe the complex plane of this device's impedance. 
The faster bolometer response requires faster excitation frequencies than we were able to generate to break degeneracies between the fitted parameters.
Thus, for this particular detector, the fitted parameters and inferred time constant is poorly constrained.
Inspecting $Z_\textrm{TES}$ of the 150 GHz Low C device confirmed this by showing that the right hand side of the semicircle in the complex plane was not measured and thus the fitted parameters, especially $\beta$, were not well constrained. 
In the future, this could be improved by using data from other measurements, such as I-V curves, to place prior constraints on $\beta$, reducing the degeneracy in the CZ fits.
Additionally, increasing the sampling and excitation rate of both the complex impedance and bias step methods would likely also improve the data and fits of our fastest devices.

\section{Conclusions}
\label{sec:conclusions}
We measured the effective thermal time constants of prototype SO MF bolometers using bias step and complex impedance methods.
The bias step method fits a single pole exponential to a bolometer's thermal response to a step in the DC bias and is useful for fast, in-situ array characterization when on-sky.
The complex impedance fits a multi-parameter bolometer model to a bolometers response to sinusoidal stimuli and while these measurements are too slow to be suitable for in-situ array-scale characterization, they reveal a broader picture of the device physics.
In both methods, the bolometer responses fit well to the single thermal block model described above.
The time constants measured in both methods are in general agreement with each other although begin to break down at faster speeds.
The complex impedance fits of fast devices poorly constrain $\beta$ because the high frequency limit of $Z_\textrm{TES}(f)$ is not measured, causing degeneracies between the fitted parameters and poor constraints on $f_\textrm{3dB}$.
The complex impedance measurements can be improved by probing at higher excitation frequencies.
Similarly, the bias step fits of the fastest devices can also be improved by sampling at higher rates in order to obtain more data on the exponential decay.
Further measurements may be explored as part of future SO single pixel characterization analyses.
The effective thermal time constants measured here are in line with the expected bolometer model and meet the SO time constant requirements.

\acknowledgements
This work was funded by the Simons Foundation (Award \#457687, B.K.).
Work by NFC was supported by a NASA Space Technology Research Fellowship.
MDN acknowledges support from NSF award AST-1454881.
SKC acknowledges support from NSF award AST-2001866.
ZX is supported by the Gordon and Betty Moore Foundation.

\bibliography{references,report} 
\bibliographystyle{spiebib} 

\end{document}